\documentstyle[12pt]{article}
\newcommand{\NN}{\nonumber}
\newcommand{\PD}[1]{\partial_{{#1}}}
\newcommand{\toda}[1]{a_{{#1}}^{(1)}}
\newcommand{\PH}{\partial}
\newcommand{\PA}{\bar{\partial}}

\newcommand{\half}{{1\over 2}}

\newcommand{\llabel}[1]{\label{#1}}

\newcommand{\HALF}[1]{\displaystyle{{#1}\over 2}}
\newcommand{\BE}{\begin{equation}}
\newcommand{\EE}{\end{equation}}
\newcommand{\BEA}{\begin{eqnarray}}
\newcommand{\EEA}{\end{eqnarray}}
\newcommand{\TMP}[3]{{\it Theor.~Math.~Phys.}~{\bf #1}~{(#2)}~{#3}}
\newcommand{\IJMP}[3]{{\it Int.~J.~Mod.~Phys.}~{\bf #1}~{(#2)}~{#3}}
\newcommand{\FAA}[3]{{\it Funct.~Analy.~Appl.}~{\bf #1}~{(#2)}~{#3}}
\newcommand{\NPB}[3]{{\it Nucl.~Phys.}~{\bf #1}~{(#2)}~{#3}}
\newcommand{\JPA}[3]{{\it J.~Phys.}~{\bf #1}~{(#2)}~{#3}}
\newcommand{\PLB}[3]{{\it Phys.~Lett.}~{\bf #1}~{(#2)}~{#3}}

\newcommand{\PTP}[3]{{\it Prog.~Theo.~Phys.}~{\bf #1}~{(#2)}~{#3}}
\newcommand{\ibid}[3]{{\it ibid.}~{\bf #1}~{(#2)}~{#3}}
\begin{document}
\begin{center}
\vspace*{1.0cm}
{\LARGE{\bf Toda Lattice Models with Boundary}}

\vskip 1.5cm

{\large {\bf Akira FUJII
\footnote{e-mail:{\sf akira@avzw03.physik.uni-bonn.de}}
}
%}

\vskip 0.5 cm

{\sl Physikalisches Institut} \\
{\sl Universit{\" a}t Bonn} \\
{\sl D53115 GERMANY.}
}

\vspace{1 cm}

\begin{abstract}
We consider the soliton solutions in 1- and (1+1)-dimensional Toda lattice
models with a boundary. We make use of the solutions already known on a full
line by means of the Hirota's method. We explicitly construct the solutions
satisfying the boundary conditions. The ${\bf Z}_{\infty}$-symmetric boundary
condition can be introduced by the two-soliton solutions naturally.
\end{abstract}
\vspace{1 cm}
\end{center}
\newpage
%\baselineskip=24pt
%%%%%%%%%%%%%%%%%%%%%%%%%%%%%%%%%%%%%%%%%%%%%%%%%%%%%%%%%%%%%%%%%%%%%%
\section{Introduction}
%%%%%%% general
Integrable models on a half line have been widely studied recently.
To investigate the integrable models on a half line, two approaches are
already known. One is the algebraic approach by the boundary Yang-Baxter
formalism\cite{cherednik} and the other is the field theoretical one
with the Lagrangian accompanied by the proper boundary
terms\cite{sklyanin}. The model
which is integrable on a full line needs the proper boundary terms
in order to be integrable also on a half line. Therefore, the
 boundary terms
should be determined by the requirement that the infinite number of
integrals of motion on a full line
must survive after introducing the boundary. Both approaches assume
that the particle content and the S-matrix are the same as
those on a full line. Though they offer quite beautiful descriptions of
the integrable models on a half line, especially in sine-Gordon model
\cite{ghoshal,grisaru} and affine Toda field theory (ATFT)
\cite{braden},
the relations between the two approaches are to be clarified.
As for the sine-Gordon
model, the semi-classical analysis has already been applied to
determine the relation of the boundary term and the boundary S-matrix
\cite{saleur}.

In the Lagrangian approach, there are two ways
to demand the infinite number of integrals of motion
to survive after introducing the boundary. In the first,
the given boundary terms determine the boundary condition of the
variables or fields through the Euler-Lagrange equation.
These boundary terms give the boundary
condition of the conserved currents, which have the same forms as
those in a full line. Therefore it is possible to define the integrals
of the conserved currents over the half line, which do not
have the asymptotic condition. If we demand those
integrals to be time-independent, we can determine the boundary terms
inversely. The second way is to use the transfer matrix.
We make the proper image of the half line at
the boundary by adding the proper boundary terms to the Lagrangian,
which enable us to consider the trace over the spatial direction.
Therefore we can define the transfer matrix on a half line properly
and demand
it to be time-independent.

%%%%%%% hirota
On the other hand, Hirota's method has been used to obtain
the soliton solutions of the equation of motion in a variety of
 integrable models on a full line in 1- or (1+1)-dimensions,
 for example, the nonlinear Schr{\" o}dinger model, sine-Gordon
model, KdV model and Toda lattice model\cite{hirota}.
This method gives the solutions in quite
an explicit form so it is an appropriate method to consider the boundary
conditions.

%%%%%%% contents
In this paper, we discuss the classical solutions of the equation
of motion with the integrable boundary conditions in 1-dimensional
Toda lattice model and the extension to the (1+1)-dimensional
field theory (Toda lattice field theory). As for the other integrable
models on a half line, the classical solutions are already known
by the {\it classical inverse scattering method}\cite{faddeev,cism}. Therefore
our analysis of Toda lattice models is considered as a first step.

%%%%%%% 1. toda lattice
This paper is organized as follows. In section 2 the Toda lattice (TL) model
is investigated.
The TL model is considered as the multi-particle systems or
the model of an electric circuit with solitons.
The equation of motion has already been solved by means of Hirota's method
\cite{hirota}.
On the other hand, the boundary terms to make the TL model
integrable on a
half line have been determined by means of the transfer matrix
method\cite{sklyanin}.
We will consider the soliton solutions obtained by Hirota's
method on a full line and satisfying the boundary condition explicitly.
We will see the meaning of the parameter in the boundary term in
several simple examples.
%%%%%%% 2. 1+1 extension
In section 3 the (1+1)-dimensional Toda lattice field theory (TLFT)
is introduced to clarify the notations.
The TLFT is an extension of the TL model to
a field theory in (1+1)-dimension and has the soliton solutions
unlike the $\toda{N}$ ATFT with a real
coupling constant.
In section 4 the TLFT with a proper boundary term
on a half line is investigated.
There seem  to be several ways to introduce the boundary in this model, for
example, the half number of the fields should vanish. However, we consider the
way to introduce the boundary in the spatial direction, which has the
connection with the $\toda{N}$ ATFT analyzed already,
therefore it is the most suitable to consider.
We will see that the boundary terms conserving the ${\bf Z}_{\infty}$-symmetry
can be induced by the two-soliton solutions.
In section 5 we present a summary of this paper.
%%%%%%%%%%%%%%%%%%%%%%%%%%%%%%%%%%%%%%%%%%%%%%%%%%%%%%%%%%%%%%%%%%%%%%
\section{Integrable Boundary Condition in Toda Lattice Model}
It is well-known that the one-dimensional TL model is
integrable, which can be shown for example
in terms of the construction by
the Lax pair.
If we consider the integrable models on a half line, integrability
imposes the restricting condition on the boundary terms, which determines
the boundary condition of the model. The integrable boundary condition
(IBC), which is compatible with the integrability, can be
 determined by the requirement that the integral of motion
on a full line must survive also on a half line. In turn, this can be
realized for example by imposing that the transfer matrix,
which is validly defined
on a half line, to be time-independent like that without the
boundary. The IBC in TL model is determined in Sklyanin's paper\cite{sklyanin}
 in this
scheme. In this section, we consider the classical solutions with the IBC
according to Hirota's method.

To begin with, we consider the TL model on a full line. The one-dimensional
TL model is defined with
the canonical coordinates $(\pi_{n},q_{n}) $, which are functions of the time
variable $t$ and on a full line, integer $n\in(-\infty,\infty)$.
The hamiltonian and the Poisson brackets are
\BE
H=\sum_{n=-\infty}^{\infty} {\pi_n ^2 \over 2}
+\sum_{n=-\infty}^{\infty} e^{q_{n+1}-q_{n}}, \qquad
\{ \pi_{m},q_{n} \} = \delta_{nm} \llabel{hamitl}.
\EE
Now we write down the equation of motion in terms of new
variables $r_{n}=q_{n+1}-q_{n}$. The equation of motion is
\BE
{\ddot r}_{n}=e^{r_{n+1}}-2e^{r_{n}}+e^{r_{n-1}}, \qquad
n=0,\pm 1,\pm 2,\pm 3,\dots, \llabel{emtl}
\EE
here \.~ represents the time derivative ${d \over dt}$.
We can obtain the soliton solutions of the equation of motion (\ref{emtl})
by means of Hirota's
method. To apply Hirota's method, we introduce the new variables
$V_{n}$ and $F_{n}$ as
$r_{n}=\ln (1+V_{n})$ and $V_{n}={d^2 \over dt^2}\ln F_{n}$. With these
variables $V_{n}$ and $F_{n}$,
the equation of motion (\ref{emtl}) can be written in a simpler form
\BE
{\ddot F}_{n} F_{n}-{\dot F}_{n}^{2}=F_{n+1}F_{n-1}-F_{n}^{2} \llabel{emtlf}.
\EE
The above equation of motion (\ref{emtlf}) has $s$-soliton solutions
\BEA
F_{n}&=&\sum_{\underline{\mu}=0,1}\exp
\left(
\sum_{i=1}^{s}\mu_{i}\eta_{i} + \sum_{i<j}A_{ij}\mu_{i}\mu_{j}
\right), \NN \\
\eta_{i}&=&p_{i}n-\Omega_{i}t+\delta_{i}, \llabel{soliton}
\EEA
where $\delta_{i}$'s are arbitrary real parameters and $\mu_{i}$
$(i=1,2,\ldots,s)$ is 0 or 1.
The real-valued momenta $(p_{i},\Omega_{i})$ and
the coefficient $A_{ij}$ must satisfy
the condition
\BEA
\Omega_{i}&=&2\epsilon_{i}\sinh{p_{i}\over 2},
\qquad \epsilon_{i}=\pm 1, \NN \\
e^{A_{ij}}&=&\left\{
\begin{array}{c}
\displaystyle{
\left(
{\sinh{1\over 4}(p_{i}-p_{j})\over\sinh{1\over 4}(p_{i}+p_{j})}
\right)^{2}} \qquad \epsilon_{i}\epsilon_{j}>0 \\
\displaystyle{
\left(
{\cosh{1\over 4}(p_{i}-p_{j})\over\cosh{1\over 4}(p_{i}+p_{j})}
\right)^{2}} \qquad \epsilon_{i}\epsilon_{j}<0
\end{array}
\right. . \llabel{tlsol}
\EEA
The above soliton solutions approach
zero as $|n|\rightarrow\infty$ and does not diverge anywhere provided
$p_{i},c_{i}$ and $\delta_{i}$ are real numbers \cite{fring,hangcho}.

We are ready to consider the TL model with the IBC.
The hamiltonian takes the form
\BE
H=\sum_{n=1}^{\infty}{\pi_{n}^{2}\over 2}+\sum_{n=1}^{\infty}
e^{q_{n+1}-q_{n}}+
\left(
\alpha e^{q_{1}}+{\beta\over 2}e^{2q_{1}}
\right) . \llabel{hamibtl}
\EE
It can be shown that the model given by the above hamiltonian is also
integrable with the arbitrary parameters $\alpha$ and $\beta$
because it has an infinite number of the integrals of
motion\cite{sklyanin}. The equation of motion can be written down with
$r_{n}=q_{n+1}-q_{n}$ ($n=1,2,\ldots$) and $q_{1}$ as
\BEA
{\ddot r}_{n}&=&e^{r_{n+1}}-2e^{r_{n}}+e^{r_{n-1}} \qquad n=2,3,\cdots,
\llabel{emrbtl0}\\
{\ddot r}_{1}&=&e^{r_{2}}-2e^{r_{1}}+(\alpha e^{q_{1}}+\beta e^{2q_{1}})
\llabel{emrbtl}.
\EEA
Eqs.(\ref{emrbtl0}) for $r_{n}$ ($n=2,3,\ldots$) are the same as those on a
full line (\ref{emtl}) and Eq.(\ref{emrbtl}) for $r_{1}$ includes the
variable $q_{1}$. We make a remark that if $\lim_{n\rightarrow\infty}
|\sum_{i=1}^{n}r_{i}|<\infty$, which is true for the soliton solution
(\ref{soliton}),
$q_{1}=(q_{1}-q_{2})+(q_{2}-q_{3})+\ldots =-\sum_{i=1}^{\infty}r_{i}$.
Comparing Eqs.(\ref{emtl}) and (\ref{emrbtl0})(\ref{emrbtl}), we obtain
the boundary condition in Eq.(\ref{emtl}), which enables the soliton solution
(\ref{soliton}) on a full line to satisfy Eqs.(\ref{emrbtl0}) and
(\ref{emrbtl}) on a half line in the form
\BE
e^{r_{0}}=\alpha\prod_{i=1}^{\infty}e^{-r_{i}}+\beta
\left( \prod_{i=1}^{\infty}e^{-r_{i}}\right)^{2}.
\EE
In terms of $F_{n}$, this can be written down simply as
\BE
F_{-1}=\alpha F_{0}+\beta F_{1} \llabel{bcftl}.
\EE
Firstly, we will concentrate the special case with
$\alpha+\beta=1, \beta\geq 0$ for simplicity.
In this case, we can find two-soliton solutions with arbitrary momentum
satisfying the boundary condition (\ref{bcftl}).
We consider the two-soliton solution
\BEA
F_{n}&=&1+e^{-\Omega t}(e^{pn+\delta_{1}}+e^{-pn+\delta_{2}})+
\left(\cosh{p\over 2}\right)^{2}e^{-2\Omega t+\delta_{1}+\delta_{2}},
\NN\\
\Omega&=&2\sinh{p\over 2},
\llabel{tl2solf}
\EEA
with an arbitrary momentum $p$.
The boundary condition(\ref{bcftl}) gives the equation of
the initial displacements $\delta_{1}$ and $\delta_{2}$
\BE
e^{\delta_{1}-\delta_{2}}={1+\beta e^{-p}\over e^{-p}+\beta}
\llabel{tlphaseshift},
\EE
which is positive-definite even if $\alpha=1-\beta$ is negative.
Therefore the system is stable\cite{fs}.
This equation means the following. We assume $\Omega>0,p>0$. In the limit
$t\rightarrow -\infty$, there exist two solitons with momenta
$(\Omega,p)$ and $(\Omega,-p)$. As the time $t$ grows to $\infty$,
$F_{n}$ goes to 1. Therefore if the initial displacements of the two
solitons, $\delta_{1}$ and $\delta_{2}$, satisfy Eq.(\ref{tlphaseshift}),
these two solitons undergo a pair-annihilation
due to the boundary. In particular, if
$\alpha=0$ and $\beta=1$, Eq.(\ref{tlphaseshift}) is simplified
as $\delta_{1}=\delta_{2}$.
It means that in this case, the shift of the initial displacements due to
the boundary
is independent of the
momentum $p$ of the soliton.

Before closing this section, we briefly mention the general case with
arbitrary $\alpha$ and $\beta$. If $\alpha+\beta\neq 1$, it is
much more difficult to find the solutions satisfying the boundary
condition (\ref{bcftl}) than those with $\alpha+\beta=1$.
We consider the case in which the initial momentum $p$ takes a special
value so that the two-soliton solutions may exist. We consider a
two-soliton solution
\BEA
F_{n}&=&1+e^{\Omega t-pn+\delta_{1}}+e^{-\Omega t-pn+\delta_{2}}
+\left(\cosh{p\over 2}\right)^{-2}e^{-2pn+\delta_{1}+\delta_{2}},
\NN\\
\Omega&=&2\sinh{p\over 2}.\llabel{neq1}
\EEA
If we assume the region of $\beta$ as $-e^{-p}<\beta<-e^{-2p}$ for $p>0$ or
$-e^{-2p}<\beta<-e^{-p}$ for $p<0$,
this soliton-solution satisfies the boundary condition (\ref{bcftl}) only if
\BEA
&& e^{-p}-\alpha+|\beta| e^{p}=0,\llabel{psn1}\\
&&e^{\delta_{1}+\delta_{2}}={|\beta|-e^{-p}\over -|\beta|
e^{p}+e^{-2p}}.\llabel{psn2}
\EEA
%%%%%%%%%%%%%%%%%%%%%%%%%%%%%%%%%%%%%%%%%%%%%%%%%%%%%%%%
where the value (\ref{psn2}) is made positive.
We assume $\Omega>0,p>0$ as before. Considering the limits $t\rightarrow\pm
\infty$, the initial soliton with momentum $(\Omega,p)$ is scattered by
the boundary to the soliton with momentum $(-\Omega,p)$ if the momentum
$p$ satisfies Eq.(\ref{psn1}) and the initial phase $\delta_{2}$
are determined from $\delta_{1}$ by Eq.(\ref{psn2}).
%%%%%%%%%%%%%%%%%%%%%%%%%%%%%%%%%%%%%%%%%%%%%%%%%%%%%%%%%%%%%%%%%%%%
\section{Definitions and Solutions in (1+1)-dimensional Toda Lattice
Field Theory}
In this section, we consider the extension of one-dimensional
TL model to a (1+1)-
dimensional field theory (TLFT).
As well as other (1+1)-dimensional integrable
classical field theories, the equation of motion of the TLFT can also
be solved by means of Hirota's method. (1+1)-dimensional TLFT
($\toda{\infty}$ Affine Toda field theory with the real coupling constant)
is defined by the action
\BE
{\cal S}=\int_{-\infty}^{\infty}d\sigma\int_{-\infty}^{\infty}d\tau {\cal L}
=\int_{-\infty}^{\infty}d\sigma\int_{-\infty}^{\infty}d\tau\left(
\half(\partial_{\tau}{\vec \phi})^{2}
-\half(\partial_{\sigma}{\vec \phi})^{2}
-\sum_{i=-\infty}^{\infty}e^{{\vec\alpha}_{i}\cdot{\vec\phi}}
\right) \llabel{actiontlft}.
\EE
Here we explain the notations briefly. Let ${\bf R}^{\infty}$ be
the infinite-dimensional vector space.
${\vec\phi}(\sigma,\tau)\in{\bf R}^{\infty}$ is an infinite-dimensional
vector field and ${\vec\alpha}_{i}={\vec e}_{i+1}-{\vec e}_{i}$, where
$\{ {\vec e}_{i} \}$ is the standard basis of ${\bf R}^{\infty}$, is a
simple root of $\toda{\infty}$ Lie algebra. If we put
${\vec \alpha}_{i}\cdot{\vec \phi}=\phi_{i}$, the equation of motion can be
written as
\BE
\PD{\tau}^{2}\phi_{n}-\PD{\sigma}^{2}\phi_{n}=
2e^{\phi_{n}}-e^{\phi_{n+1}}-e^{\phi_{n-1}},
\qquad n=0,\pm 1,\pm 2,\cdots . \llabel{emtlftphi}
\EE
We carry out the calculation in the light-cone
coordinate $z=(\sigma-\tau)/2$ and $\bar{z}=(\sigma+\tau)/2$
($\PH=\PD{z},\PA=\PD{\bar{z}}$) to keep the notation simple.
If we introduce  new variables $F_{n}$ and $V_{n}$
$\phi_{n}=\ln(1+V_{n})$ and $V_{n}=-\PH\PA\ln F_{n}$ as before,
the equation of motion of the model (\ref{actiontlft}) can be
written in terms of $F_{n}$'s as
\begin{equation}
\PH F_{n}\PA F_{n}-F_{n}\PH\PA F_{n}=
F_{n+1} F_{n-1}-F_{n}^{2} \llabel{tlftemf}.
\end{equation}
The equation of motion (\ref{tlftemf}) has the $s$-soliton solutions
\BEA
F_{n}&=&\sum_{{\underline \mu}=0,1}\exp\left(\sum_{i=1}^{s}
\mu_{i}\rho_{i}
+\sum_{i<j}A_{ij}\mu_{i}\mu_{j}\right) ,\NN\\
\rho_{i}&=&a_{i}z+b_{i}{\bar z}+c_{i}n+\delta_{i}, \llabel{tlftsolf0}
\EEA
here the initial phases $\delta_{i}$ are arbitrary and the momenta
$a_{i}$,$b_{i}$ and $c_{i}$ and the coefficients $A_{ij}$ must satisfy the
condition
\BEA
a_{i}b_{i}&=&-4\sinh^{2}\HALF{c_{i}}, \NN\\
e^{A_{ij}}&=&-{
(a_{i}-a_{j})(b_{i}-b_{j})+4\sinh^{2}\HALF{c_{i}-c_{j}}
\over
(a_{i}+a_{j})(b_{i}+b_{j})+4\sinh^{2}\HALF{c_{i}+c_{j}}
} \llabel{tlftsolf}.
\EEA
Several remarks are in order. Firstly,
from the requirement that the solutions given by (\ref{tlftemf}) do not
diverge on a full line, {\it i.e.} $V_{n}>-1$, we will get the
condition that $e^{A_{ij}}\geq 0$
and that $\delta_{i}$'s are real values. Secondly, the
above given solutions are soliton solutions.
It is well-known that contrary to the sine-Gordon case, there is no soliton
solutions in the (real-coupling) $\toda{N}$ ATFT for
finite $N$, for example sinh-Gordon model.
The reason is that in the case of $\toda{N}$ ATFT for the finite $N$,
the cyclic condition in the direction of the affine coordinate, that is
$i+N\equiv i$, demands the momenta $c_{i}$ to take certain complex
values $c_{i}=(2\pi i/N){\bf Z}$. This fact causes the naive
soliton solutions given in Eq.(\ref{tlftsolf0}) to diverge to infinity
in the limit $\sigma\rightarrow\pm\infty$, therefore these models do not
have soliton solutions.
The TLFT, in spite of corresponding to the
$\toda{\infty}$ ATFT with the real coupling constant,
has no cyclic condition in the affine direction
so we can choose $c_{i}$ to be real values, which do not cause the
divergence.
Conversely, we cannot obtain the solutions in TLFT by merely
taking the limit of
$N\rightarrow\infty$ in $\toda{N}$ ATFT. We need to take the analytic
continuation in the affine direction.
%%%%%%%%%%%%%%%%%%%%%%%%%%%%%%%%%%%%%%%%%%%%%%%
\section{Toda Lattice Field Theory with Boundary}
In this section, we will consider the TLFT on a half line. We choose
the space coordinate $\sigma$ in $(-\infty,0)$
and the time coordinate $\tau$ in $(-\infty,\infty)$. The action of the
TLFT on a half line is
given by
\BE
{\cal S}=\int_{-\infty}^{0}d\sigma\int_{-\infty}^{\infty}d\tau {\cal L}
+\int_{-\infty}^{\infty}d\tau{\cal L}_{B} \llabel{actionbtlft},
\EE
here the {\it bulk} Lagrangian density ${\cal L}$ is
the same as that given before on
a full line (\ref{actiontlft})
and the boundary term ${\cal L}_{B}$ is given as
\BE
{\cal L}_{B}=-\sum_{i=-\infty}^{\infty} {\cal A}_{i}
e^{{\vec\alpha}_{i}\cdot{\vec\phi}/2}\bigl|_{\sigma=0}
 \llabel{bactiontlft},
\EE
with the parameters ${\cal A}_{i}$.
It is well-known that the $\toda{N}$-ATFT with a finite $N\geq 2$ defined
by the above action (\ref{bactiontlft})
is integrable if $|{\cal A}_{i}|=0$ or 2 for all $i$\cite{braden}.
In these cases, it is shown
that the integrals of motion on a full line can survive after introducing
the boundary at $\sigma=0$.
Now we consider the classical (soliton) solutions in this model
on a half line. The equation of motion is the same as that
on a full line (\ref{emtlftphi}) and the boundary term ${\cal L_{B}}$ in
the action (\ref{actionbtlft}) induces the boundary condition of ${\vec \phi}$
at $\sigma=0$
\BE
\PD{\sigma}\phi_{n}\left|_{\sigma=0}=
2{\cal A}_{n}e^{\phi_{n}/2}-{\cal A}_{n+1}e^{\phi_{n+1}/2}-{\cal
A}_{n-1}e^{\phi_{n-1}/2}
\right. . \llabel{bctlftphi}
\EE
For the later convenience, we write down the boundary condition
(\ref{bctlftphi}) in terms of the variable $F_{n}$, which is defined the
same as that on a full line,
\begin{equation}
\left(2{F_{n}'\over F_{n}}-{F_{n+1}'\over F_{n+1}}-{F_{n-1}'\over F_{n-1}}
\right)
\bigl|_{\sigma=0}=\left(
{\cal A}_{n}{\sqrt{F_{n+1}F_{n-1}}\over F_{n}}-
{{\cal A}_{n+1}\over 2}{\sqrt{F_{n+2}F_{n}}\over F_{n+1}}-
{{\cal A}_{n-1}\over 2}{\sqrt{F_{n-2}F_{n}}\over F_{n-1}}
\right)\bigl|_{\sigma=0}
\llabel{bctlftf},
\end{equation}
here $'$ means the spatial derivative $\partial/\partial\sigma$.
We will consider the soliton solutions (\ref{tlftsolf0})
obtained on a full
line, some of which satisfy the boundary condition (\ref{bctlftf}).

Firstly,
we consider the simplest example; ${\cal A}_{i}=0$ for all $i$.
If a solution $F_{n}$ is an even function
in $\sigma$, it satisfies the boundary condition (\ref{bctlftf}).
In particular, if we consider two-soliton solutions ($s=2$)
in (\ref{tlftsolf0})
with
\begin{equation}
\rho_{1}=p\sigma-\epsilon\tau +cn+\delta, \qquad
\rho_{2}=-p\sigma-\epsilon\tau +cn+\delta,
\end{equation}
the coefficient $A_{12}$ takes quite a simple form as
\BE
e^{A_{12}}=\left(1-\left({2\over
p}\sinh^{2}\HALF{c}\right)^{2}\right)^{-1}\llabel{eatl},
\EE
from the condition (\ref{tlftsolf}).
Providing this solution does not diverge on a half line, the coefficient
(\ref{eatl}) must be positive, that is $|p|\geq 2\sinh^{2}{c\over 2}$.
%%%%%%%%%%%%%%%%%%%%%%%%%%%%%%%%%%%%%%%%%%%%%%%%%%%%%%%%%%%%%%%%%%%%%%%%%

Secondly,
we consider the general boundary condition given by (\ref{bctlftphi}).
We assume two-soliton solutions ($s=2$) given by (\ref{tlftsolf0}).
Because the lefthand-side of Eq.(\ref{bctlftf})
is a rational function of $e^{\rho_{1}}$ and $e^{\rho_{2}}$,
we must impose that the square root
$\sqrt{F_{n+1}F_{n-1}}|_{\sigma=0}$ takes the form of a certain polynomial in
$e^{\rho_{1}}$ and $e^{\rho_{2}}$.
This condition makes the
calculation much
simpler and gives the constraints of the momenta as
\BEA
&&\rho_{1}=p\sigma-\epsilon\tau+cn+\delta_{1},\qquad
\rho_{2}=p\sigma+\epsilon\tau-cn+\delta_{2}, \NN \\
&& e^{A_{12}}=1-\left({2\over p}\sinh^{2}{c\over 2}\right)^{2} , \llabel{zn}
\EEA
where momenta $p,\epsilon$ and $c$ are arbitrary real values with the
condition (\ref{tlftsolf}) and the phase displacements $\delta_{1}$ and
$\delta_{2}$ must satisfy the equation
\BE
e^{\delta_{1}+\delta_{2}}={1\over 1+\gamma\left({2\over p}\sinh^{2}{c\over
2}\right)}
,
\qquad \gamma=\pm 1. \llabel{znps}
\EE
Due to the condition of the absence of the divergence, {\it i.e.}
$V_{n}>-1$, we must insist $|p|\geq 2\sinh^{2}{c\over 2}$ as before.
With the above preparation, we can consider the boundary condition
(\ref{bctlftf}). Because we can easily obtain
\BE
\left(F'_{n}+\gamma \sqrt{F_{n+1}F_{n-1}}-(p-\gamma)F_{n}\right)
\bigl|_{\sigma=0}=0, \llabel{znsa}
\EE
we see that ${\cal A}_{i}=2\gamma$ for all $i$.
Therefore, if we consider the two-soliton solutions, we automatically
obtain the ${\bf Z}_{\infty}$-preserving boundary condition,
${\cal A}_{i}=2$ or $-2$ for all i.

We note that if only the soliton solutions are considered,
it would be quite difficult to find the solutions satisfying the
${\bf Z}_{\infty}$-symmetry breaking boundary condition, for example
${\cal A}_{0}=-2$ and ${\cal A}_{i}=2$ for $i\neq 0$.
In particular, even if we introduce the three-soliton solutions with
\BEA
&&\rho_{1}=p\sigma-\epsilon\tau+cn+\delta_{1},\quad
\rho_{2}=p\sigma+\epsilon\tau-cn+\delta_{2},\NN \\
&&\rho_{3}=\pm 2\sigma+i\pi n+\delta_{3}, \llabel{3so}
\EEA
which is the unique choice for the three-soliton solution to
satisfy the boundary condition(\ref{bctlftphi}) and to have no divergence,
we obtain the same result as that in the case of the two-soliton
solutions (\ref{znsa}). It means that the third soliton given by (\ref{3so})
has no effect on the boundary condition.
%%%%%%%%%%%%%%%%%%%%%%%%%%%%%%%%%%%%%%%%%%%%%%%%%%%%%%%%%%%%%%%%%%%%%%%%%%
\section{Summary}
In this paper, we have considered the TL model and TLFT with boundary
and seen the meaning of the parameters in the boundary term.
A few comments are in order. Firstly, the analysis in this paper can
be applied only to the other models with the soliton solutions.
Therefore, it is impossible to consider the $\toda{N}$ ATFT
with finite $N$ and real coupling constant,
which do not have the soliton solutions in the same way.
However, it is possible to consider the $\toda{N}$ ATFT
with imaginary coupling constant. Especially, the analysis of TLFT with
the Neumann-type boundary condition can be similarly used in that for
$\toda{N}$ ATFT with imaginary coupling constant.
Secondly, we did not analyze TL with
arbitrary $\alpha$ and
$\beta$ nor TLFT with general ${\cal A}_{i}$'s.
To analyze these models, we must take multi-soliton solutions into account.

{\bf Acknowledgement}

The author is grateful to R.~Sasaki for critical reading the manuscript and
discussion. He also acknowledges
G.~Albertini and V.~Rittenberg for reading the manuscript. This work
is financially supported by the AvH Foundation.
%%%%%%%%%%%%%%%%%%%%%%%%%%%%%%%%%%%%%%%%%%%%%%%%%%%%%%%%%%%%%%%%%%%%%%%%
%%%%% references %%%%%

\end{document}